\documentclass[
               screen, 
                ]{acmart}

\usepackage{enumitem}
\usepackage{listings}
\usepackage{makecell}
\usepackage{blindtext}
\usepackage{textcomp}
\usepackage{balance}
\usepackage{soul}
\usepackage{graphicx}
\usepackage{hyperref}
\hypersetup{
    colorlinks=true,
    linkcolor=blue,
    filecolor=magenta,      
    urlcolor=cyan,
    pdftitle={Overleaf Example},
    pdfpagemode=FullScreen,
    }

\usepackage{xcolor}
\usepackage{colortbl}
\definecolor{light-blue}{RGB}{150, 200, 250}
\definecolor{very-light-blue}{RGB}{215, 230, 250}
\sethlcolor{yellow}

\setcopyright{acmcopyright}
\copyrightyear{2023}
\acmYear{2023}
\acmDOI{XXXXXXX.XXXXXXX}

\acmConference[Koli Calling '23]{}{November 16--19, 2023}{Koli, Finland}
\begin{document}

\title{What Skills Do You Need When Developing Software Using ChatGPT? (Discussion Paper)}

\author{Johan Jeuring}
\affiliation{%
   \institution{Utrecht University and National Education Lab Artificial Intelligence}
   \country{The Netherlands}}
\email{j.t.jeuring@uu.nl}

\author{Roel Groot}
\affiliation{%
   \institution{Utrecht University}
   \country{The Netherlands}}
\email{grootroel@gmail.com}

\author{Hieke Keuning}
\affiliation{%
   \institution{Utrecht University}
   \country{The Netherlands}}
\email{h.w.keuning@uu.nl}

\begin{abstract}
Since the release of LLM-based tools such as GitHub Copilot and ChatGPT the media and popular scientific literature, but also journals such as the Communications of the ACM, have been flooded with opinions how these tools will change programming. The opinions range from ``machines will program themselves'', to ``AI does not help programmers''. Of course, these statements are meant to to stir up a discussion, and should be taken with a grain of salt, but we argue that such unfounded statements are potentially harmful. Instead, we propose to investigate which skills are required to develop software using LLM-based tools. 

In this paper we report on an experiment in which we explore if Computational Thinking (CT) skills predict the ability to develop software using LLM-based tools. Our results show that the ability to develop software using LLM-based tools can indeed be predicted by the score on a CT assessment. There are many limitations to our experiment, and this paper is also a call to discuss how to approach, preferably experimentally, the question of which skills are required to develop software using LLM-based tools. We propose to rephrase this question to include \textit{by what kind of people/programmers}, to develop \textit{what kind of software} using \textit{what kind of LLM-based tools}. 
\end{abstract}



\maketitle

\section{Introduction}

LLM-based tools such as GitHub Copilot and ChatGPT have taken computing education (and the rest of the world) by storm, and already now perform at the level of good CS1/CS2 students~\cite{finnie2022robots,finnie2023my,prather2023robots}. With the current speed of development, these tools are expected to become even better in the not too distant future. In a recent opinion piece in the Communications of the ACM (CACM), Welsh even goes as far as saying that ``It seems totally obvious to me that of course all programs in the future will ultimately be written by AIs, with humans relegated to, at best, a supervisory role''~\cite{welsh2022end}. This is not an original opinion: already in 1961, Peter Elias, then the chair of Electrical Engineering at MIT, suggested that programming would one day be unnecessary, because the computer would find out what the programmer wanted~(\cite{greenberger1962computers}, quoted in \cite{guzdial2015learner}). When the deep learning based programming synthesis tool Deepcoder was released~\cite{balog2016deepcoder}, the New Scientist interviewed MIT professor Armando Solar-Lezama, who said ``This kind of approach could make it much easier for people to build simple programs without knowing how to write code...''~\cite{reynolds2017ai}. These statements are not only made in the scientific literature. For example, Farhad Manjoo wrote an opinion piece entitled ``It’s the End of Computer Programming as We Know It. (And I Feel Fine.)'' in the New York Times on June 2, 2023, and the educational psychologist Paul Kirschner says that ``Given the development of machine learning, I wouldn't be surprised if machines would program themselves'' in a Dutch magazine for school leaders~\cite{kader}. 

Has Elias' ``one day'' arrived yet, after more than 60 years? When looking at the current practice: obviously not. As a response to Welsh' opinion, Grady Booch tweeted\footnote{Grady Booch, Twitter, January 1, 2023}: 
\begin{quote}
Programming will be obsolete. 
-- Matt Welsh 

Nope. The entire history of software engineering is one of rising levels of abstraction
-- Me
\end{quote}
In another response, Yellin~\cite{yellin2023premature} extensively argues why deep programming (programming using deep learning approaches) won't replace programming. One of his reasons is that someone needs to specify what software they want, describe the requirements the software needs to satisfy, and check that the generated software fulfills the requirements. Also in CACM,  Meyer~\cite{meyer} discusses why a ``sloppy assistant'' isn't going to help a software developer.

Predicting the future is hard, of course, but we argue that statements about machines programming themselves are misleading. Programming corresponds to solving real-world problems (of any kind)~\cite{lonati2022we}, and AI that can solve those problems is far away. Yellin~\cite{yellin2023premature} and Meyer~\cite{meyer} convincingly show examples where AI cannot, or only partially, help in solving programming problems. On the other hand, the developments in AI support for programming are very likely going to change software development to a certain extent. We see productivity increasing in some cases when programmers use LLM-based tools~\cite{peng2023impact}, and a recent survey\footnote{\url{https://github.blog/2023-06-13-survey-reveals-ais-impact-on-the-developer-experience/}} from a company providing AI tools for software development showed that more than 90\% of 500 US-based developers used AI tools in their software development work.

Assuming that the required competencies of people that develop software need to change, how should they change? What competencies are needed for ``playing the supervisory role''~\cite{welsh2022end} when conversing with LLM-based tools, if not programming itself? We already know quite a bit about the nature of programming expertise~\cite{parnin2017nature}; does that change in the context of LLMs?

Computational thinking (CT) has been introduced as ``a way humans solve problems with the help of a computer''~\cite{wing2006computational}. Wing characterizes it amongst others by ``Conceptualizing, not programming''. Various authors have proposed more detailed definitions for CT skills. Selby and Woollard~\cite{selby2013computational} combined the results of a number of these studies and propose that CT consists of the ability to think in abstraction, decomposition, algorithms, evaluation, and generalization. Computational thinking is the problem-solving process that can lead to code, programming is writing the actual code. So if programming itself is not required anymore for developing software with the help of LLM-based tools, because the LLM-based tools write the code, maybe CT skills are?

In this discussion paper we report on an experiment in which we try to determine if a higher score on a CT skills assessment predicts better skills in developing software using LLM-based tools. In the experiment we first test a participant's CT skills, and then determine how well the participant can develop software using LLM-based tools. Rather unsurprisingly, we indeed found that CT skills predict software development skills using LLM-based tools, but the experiment raises more questions than it answers. In this paper we want to discuss what steps we should take from here to find out more about the competencies required for developing software using LLM-based tools.

Our work is related to the recent research of Kazemitabaar et al.~\cite{kazemitabaar2023studying}, who show that novices (age 10 -- 17) benefit more from using LLM-based tools in a Python programming course when they had prior programming experience in Scratch. Interestingly, but a slightly circular argument in the context of this paper, other research shows that using ChatGPT to solve programming tasks is a good way to improve CT skills~\cite{yilmaz2023effect}. In other related work, Denny et al.~\cite{denny2023promptly} develop an approach and a learning environment in which students learn how to write prompts for LLMs, and Babe et al.~\cite{babe2023studenteval} analyse the quality of prompts written by students who have only completed one programming course. 

This paper is organised as follows. In Section~\ref{sec:method} we describe our research question, and the method we use to answer our question. Section~\ref{sec:results} describes our results. Section~\ref{sec:discussion} discusses the results and their limitations, and several ideas about how to make further progress on answering our question. Section~\ref{sec:conclusions} concludes.

\section{Method}
\label{sec:method}
The research question we investigate in this paper is:
\begin{quote}
    Does the score on a CT skills assessment predict skills in developing software using LLM-based tools?
\end{quote}
We answer this question by performing an experiment. In the experiment we first determine CT skills of a participant, and then let the participant develop software using ChatGPT. In the rest of this section we describe the instruments, and the setup of the experiment. 

\subsection{Determining CT skills}
\begin{figure}
    \centering
    \includegraphics[scale=0.4]{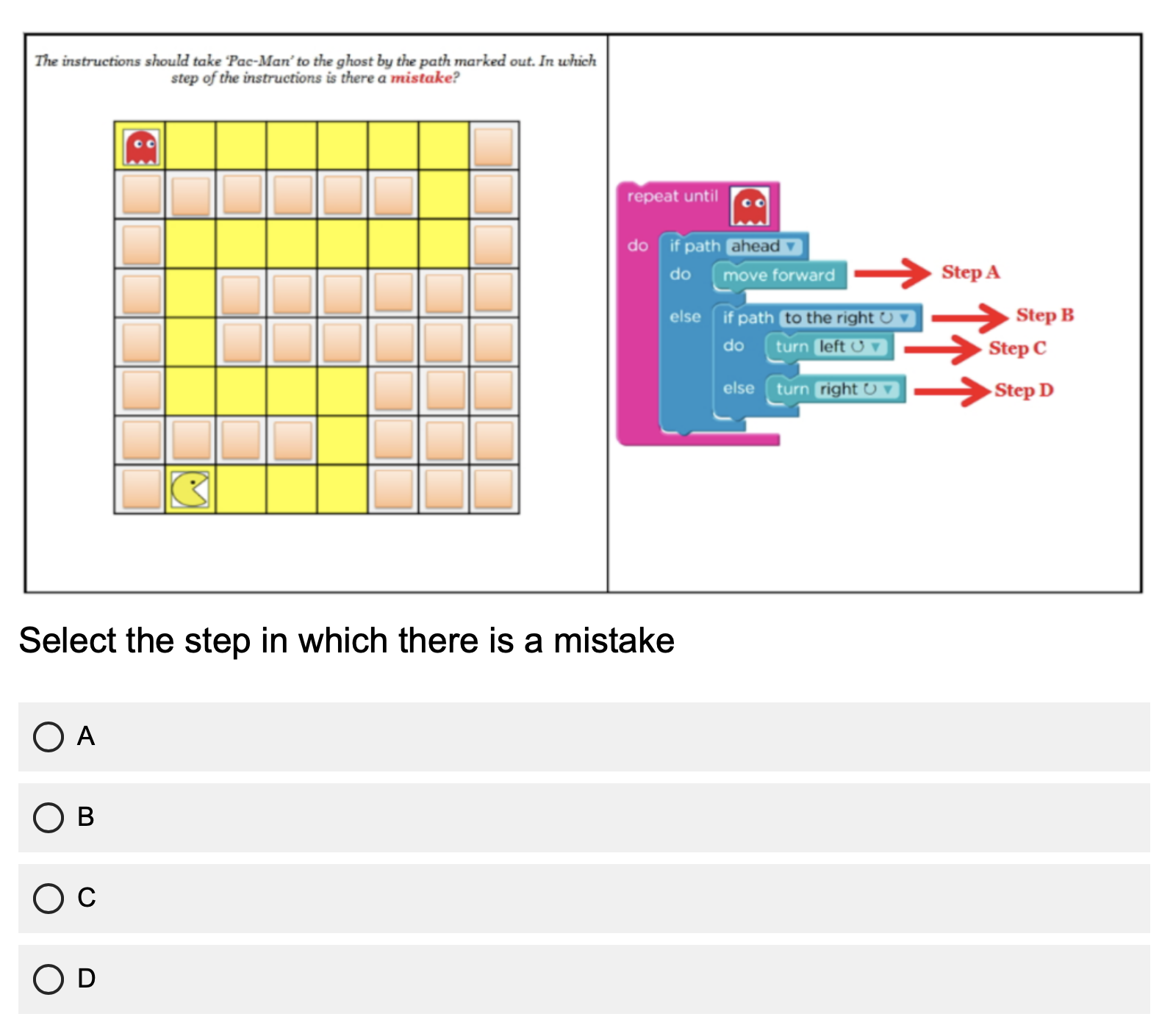}
    \caption{One of the items used in the lean computational thinking abilities assessment for middle grades students developed by Wiebe et al.~\cite{wiebe2019development}. This item is originally from the CTt test from Román-González et al~\cite{roman2018can}. Published with permission from the author.}
    \label{fig:coreCSquestion}
\end{figure}
To determine the CT skills of a participant without requiring programming skills, we need an assessment of CT skills. Since we wanted our experiment to last no longer than 45 minutes, we searched for an assessment that takes less than half an hour to complete. There are surprisingly few such tests available~\cite{tang2020assessing}. We selected the lean computational thinking abilities assessment for middle grades students developed by Wiebe et al.~\cite{wiebe2019development}. An example of an assessment question can be found in Fig.~\ref{fig:coreCSquestion}. Although seven items use block-based languages concepts, prior knowledge about these languages didn't seem to make a difference.
We selected this test because there is evidence for its validity and reliability, be it for children of age 11 -- 13, and because its duration suited the length of our experiment. 

The test consists of 23 multiple-choice items, and we scored the test by counting the number of correct answers. 

\subsection{Developing software using ChatGPT}
\begin{figure}
\begin{center}
    \includegraphics[scale=0.5]{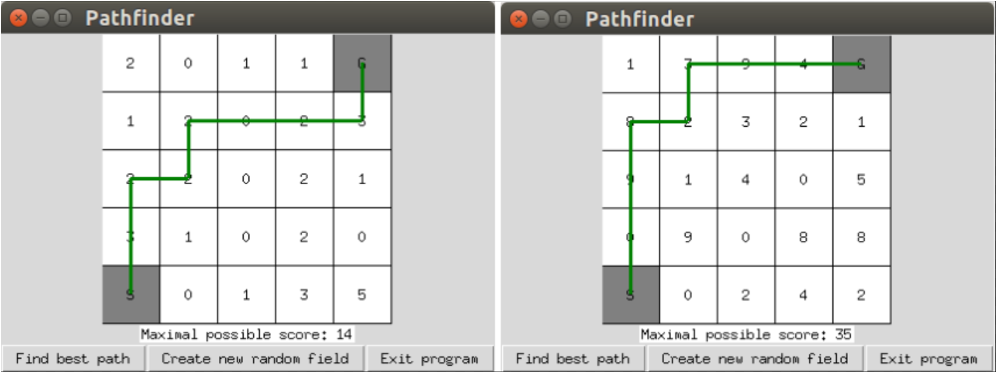}
\end{center}
\caption{Visual representations of the path.}
\label{figuretask}
\end{figure}

We asked the participants to solve a programming assignment by asking ChatGPT to generate the code. Participants could not edit the code themselves. We wanted an assignment with the following characteristics:
\begin{itemize}
    \item it shouldn't be too simple: ChatGPT shouldn't immediately solve it correctly, and it should take somebody with good programming skills more than 10 minutes to solve without using ChatGPT;
    \item it shouldn't be too hard: it should be possible for a participant with good programming skills to solve the assignment in approximately 20 minutes using ChatGPT;
    \item it shouldn't spell out all requirements of the assignment explicitly, so that participants can show that they can deal with incomplete, ambiguous, or sometimes even conflicting requirements;
    \item the task should be at the level of a CS1 final exam.
\end{itemize}
We selected a task offered in the introductory course Programming with Python at Utrecht University in 2021. We rewrote the task to not use programming language specific terminology, so that participants unfamiliar with programming could also understand the task prompt: 
\begin{quote}\textit{
Your task is to create a small program called pathfinder with the programming language Python. Please ask ChatGPT to use Python as its programming language.
This program generates a grid of random numbers, finds a path with the highest score, moving only up or right, and draws it. See Fig~\ref{figuretask} for visual representations\footnote{Distributed by Anna-Lena Lamprecht, see \url{https://github.com/annalenalamprecht/CoTaPP},
under the {Creative Commons Attribution 4.0 International License}.} of two example grids, and the paths your program should find. }
\end{quote}

We used the Legacy 3.5 version of ChatGPT. Participants only needed a computer and a stable internet connection to participate in the experiment.
We also provided some information about testing code: 
\begin{quote}\textit{
To test the code you write with ChatGPT, go to \url{https://onecompiler.com/python}, paste your code in the editor available on that web page, and press the RUN button in the top right, or press CTRL + ENTER. Alternatively, you can ask ChatGPT itself for an example of what the code does.}
\end{quote}
The online compiler can only be used to test functionality; it is not possible to test a user interface in it.
 
From the task description we distilled 13 requirements: 
\begin{enumerate}[align=parleft,leftmargin=5\parindent]
    \item[\textbf{req1}] The grid contains 5 rows and 5 columns.
    \item[\textbf{req2}] The elements of the grid contain random numbers between 0 and 9.
    \item[\textbf{req3}] The grid contains an ``S'' in the bottom left, and a ``G'' in the top right.
    \item[\textbf{req4}] The program keeps track of a path and its elements.
    \item[\textbf{req5}] The path starts in the bottom left corner and ends in the top right corner.
    \item[\textbf{req6}] The path can only move up or right, never left or down.
    \item[\textbf{req7}] The score of a path is the sum of its elements.
    \item[\textbf{req8}] The program uses an algorithm to find the path with the highest possible score.
    \item[\textbf{req9}] The program outputs the maximum score of a path it has found.
    \item[\textbf{req10}] The program visualizes the path.
    \item[\textbf{req11}] The program contains a button that
    \begin{enumerate}
        \item finds the best path and draws it;
        \item creates a new 5x5 grid with random numbers;
        \item exits the program.
    \end{enumerate}
\end{enumerate}
Implementing a requirement results in a score of 1. Since the path with the highest score was the central requirement (\textbf{req8}), and maybe also one of the more challenging ones, we awarded 2 extra points for successfully completing this requirement. All solutions were independently graded by two of the three authors, and differences in grading were discussed and resolved. 

Besides scoring the implemented requirements of the task, we also looked at the ChatGPT prompts the participants wrote. We distinguished two cases: a good prompt asked for one or more requirements to be implemented or to provide an alternative solution; a bad prompt showed that a participant was confused, misunderstood ChatGPT's remarks, or misunderstood the task. All prompts were independently assessed by two of the three authors, and differences in assessment were discussed and resolved. Some participants asked for an implementation of multiple requirements in a single prompt, where other participants asked ChatGPT to implement requirements one by one. We did not want to award different scores to these approaches. We decided to deduct 1 point from the total score for the programming task for every bad prompt. So the final scoring rule for the programming task \textit{PS} we used was:
\[ \textit{PS} = \textit{Reqs} + \textit{if \textbf{req8} then 2} - \textit{BadPrompts} \]

\subsection{The experiment}
The Ethics and Privacy Quick Scan of Utrecht University
classified this research as low-risk with no fuller ethics review or privacy assessment required. 

We recruited 19 participants through convenience sampling. We selected a mixed group of people with and without programming experience, between approximately 20 and 60 years old, and of different genders. Most participants had either completed a higher education degree or were higher education students. The participants with programming experience had a variety of programming skill levels. All participants had sufficient understanding of what a computer does and how to operate one. All participants had a good understanding of English, and were spread throughout Europe (6 different countries) with approximately half from the Netherlands.

The experiment took place in March 2023.
One of the authors supervised the experiment with all participants individually online. The participants were informed about the research and agreed to participate. They then worked on the online CT assessment for 20 minutes. After 20 minutes the CT assessment was concluded, and after a short instruction on how ChatGPT can be used to develop software, we gave the participants the programming assignment to be solved using ChatGPT, together with a ChatGPT account with no conversation history. After 18 minutes the participants would be told that there were 2 minutes left, and that they should finalize and submit their solution to the task. 

\section{Results}
\label{sec:results}
Table~\ref{tab:my_label} shows the result of the experiment. 

\begin{table}[h]
    \centering,
    \caption{Scores for each participant. \textit{CTS} is the score on the CT assessment.}
\begin{tabular}{l|rrrrrrrrrrrrrrrrrrr}

Participant nr	& 1	&2	&3	&4	&5	&6	&7	&8	&9	&10	&11	&12	&13&	14	&15	&16	&17	&18&	19\\	\hline												
\textit{CTS}&	22&	16&	17&	17&	17&	17&	21&	17&	15&	17&	9&	20&	16&	9&	9&	12&	14&	13&	11\\
\textbf{req8}&	2&	2&	2&	0&	0&	0&	0&	0&	0&	0&	0&	0&	0&	0&	0&	0&	0&	0&	0\\
\textit{Reqs}&	10&	3&	6&	4&	6&	7&	5&	8&	3&	7&	0&	7&	6&	0&	0&	5&	5&	2&	6\\
\textit{BadPrompts}&	0&	3&	0&	0&	0&	0&	0&	1&	1&	0&	10&	0&	0&	0&	5&	0&	0&	1&	2\\
\textit{PS}	&13&	3&	9&	4&	6&	7&	5&	7&	2&	7&	-10&	7&	6&	0&	-5&	5&	5&	1&	4

    \end{tabular}
     
    \label{tab:my_label}
\end{table}
\noindent
Both the CT skills scores and the scores on the programming task are normally distributed, with Shapiro-Wilk Sig.~values of 0.175 and 0.272, respectively.

\subsection{Implemented requirements}
\begin{figure}
    \centering
    \includegraphics[width=10cm]{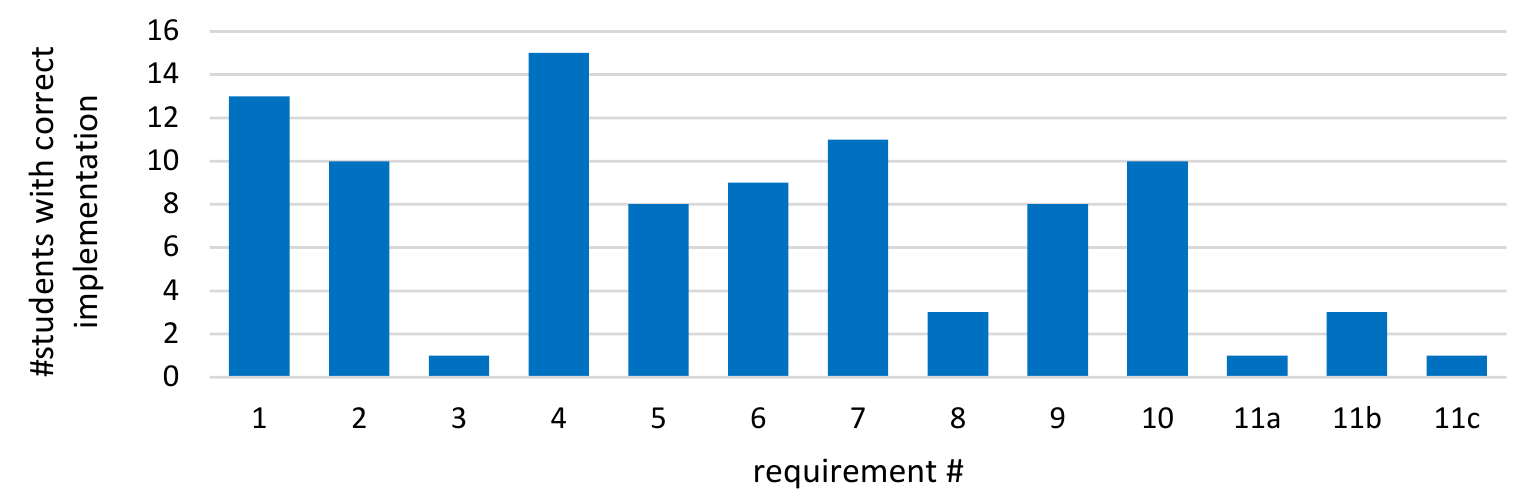}
    \caption{Number of correct implementations per requirement.}
    \label{fig:frequency}
\end{figure}

Fig~\ref{fig:frequency} shows the implementation frequency of the various requirements of the programming task. Requirements \textbf{req3} (the grid contains an ``S'' in the bottom left, and a ``G'' in the top right), \textbf{req8} (the core algorithm), \textbf{req11}.a, \textbf{req11}.b and \textbf{req11}.c (constructing the three buttons at the bottom of the application) were the least implemented requirements of all. We expected that  \textbf{req8} would be the hardest to implement correctly, since this required more algorithmic thinking than the others. The lack of implementations of \textbf{req3} is a bit harder to understand, as the picture illustrating the program shows it. We checked the prompts, and hardly any participant asked about this requirement. We speculate that the picture wasn't clear enough about this aspect. In addition, many participants didn't ask ChatGPT to implement the buttons (\textbf{req11}). This was maybe also caused by the suggestion to test code using an online compiler that could not deal with user interfaces. So participants couldn't test the generated user interface with the suggested compiler. We noticed multiple times that ChatGPT suggested code that implemented requirements without the participant asking for it, and did not implement requirements despite the participant asking for it. 

\subsection{What prompts did participants write?}
We checked the prompt the participants wrote, and assessed their quality. As mentioned in Section~\ref{sec:method}, we deducted one point from the programming score for every prompt we assessed as bad. In this section we give some examples of both good and bad prompts. 

None of the three participants scoring at least 20 on the CT skills test wrote a bad prompt. Besides asking for implementations of the requirements formulated in the text and accompanying picture of the task, they would also ask things like: ``The last lines seem to be cut off. Could you please fill those in, in order to achieve what was previously mentioned?'', and ``Please remove the input handling, and make the program simply calculate and render the path for the generated board''. 

Bad prompts varied quite a bit, and we failed to understand the underlying reasoning for most of them. For example, one of the three participants scoring zero on the programming task wrote the prompt ``graph = A: set([0-5]\}'', and another, after asking for a grid of 5x5, ``Create a line that goes from the bottom 5 up 2''. One pattern we did notice is that some participants asked ChatGPT to solve the problem described in the error message of the online Python interpreter they received when testing their code. Some of these error messages were clearly beside the point, but participants got hung up on them anyway. For example, one participant asked ChatGPT ``set the MPLCONFIGDIR environment variable to a writable directory'' multiple times in various forms. Often bad prompts led to a sequence of interactions that brought the participant further away from a good solution.

\subsection{Analysis}
The scores on the implemented requirements and bad prompts were negatively correlated, with $1/Slope$ equal to $-2.0$. This justifies combining these scores in a single score for programming skills using LLM-based tools.

\begin{figure}
    \centering
    \includegraphics[scale=0.4]{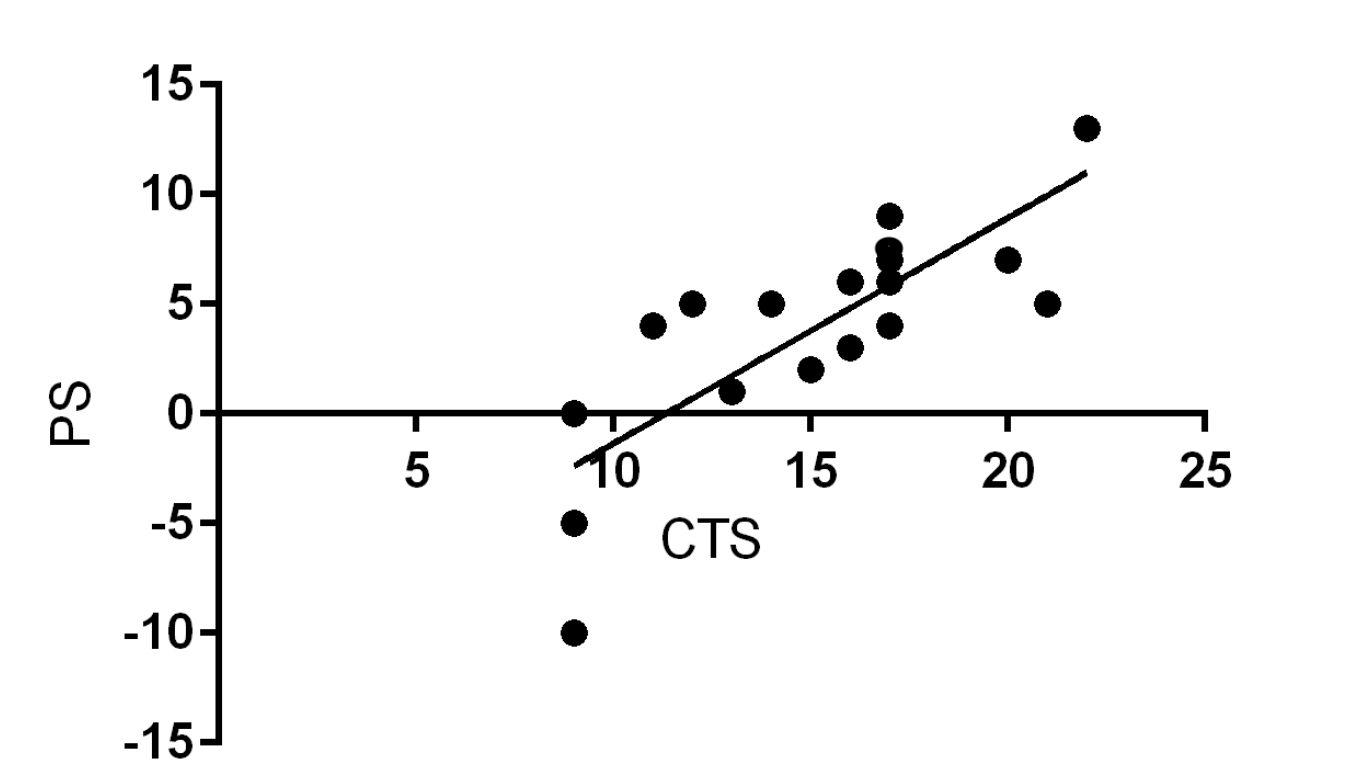}
    \caption{Scatter plot of \textit{CTS} versus \textit{PS} scores}
    \label{fig:scatterplot}
\end{figure}

To show whether or not CT skills ($CTS$) can explain programming skills using LLM-based tools ($PS$), we created a scatter plot, see Fig.~\ref{fig:scatterplot}. Note that (17,7) appears three times in the data, which isn't visible very well in the scatter plot. 
The scatter plot shows that there might be a correlation between computational thinking skills and programming with LLM-based tools. To find out whether or not such a correlation exists, we use a bivariate regression analysis, with the computational thinking skills as predictor. 
The regression analysis gives the formula 
\[ \textit{PS} = 1.029*\textit{CTS} - 11.65\] 
with a $95\%$ confidence interval $[0.6239,1.434]$ for the $Slope$. The goodness of fit value $R^{2}$ is $.63$. The $Slope$ is significantly non-zero $(F=28.75;DFn,DFd=1,17;p<0.0001)$. This implies that the computational thinking score predicts the programming with LLM-based tools score: every extra point on the CT assessment leads to (approximately) an extra point on the programming skills with LLM-based tools test. 

\section{Discussion}
\label{sec:discussion}
The many statements of the form ``with the availability of advanced LLM-based tools such as ChatGPT, people don't need to know how to program anymore'' inspired us to perform the experiment described in this paper. These statements are often not very nuanced, of course, and might mean several things, ranging from it is not necessary anymore to master the syntax of a particular programming language, to having no clue about central computer science concepts, such as file, program, user-interface, compiler, etc. Of course people need some skills to use LLM-based tools, if only to express the functionality of the desired software in language.  

To determine what skills people need to develop software using LLM-based tools is not easy. Since the description of CT skills is closest to the kind of skills we expect from somebody solving problems using a computer, we chose to determine if CT skills could predict skills in developing software using LLM-based tools. Our results show that CT skills indeed predict programming skills with LLM-based tools for people at the higher education level. 


This result of course raises a number of other questions: what skills are needed for developing CT skills, and how can we develop CT skills? By now, many interventions to develop CT skills in different subjects, mainly in STEM, have been designed~\cite{tang2020assessing}. Some of these show good results. The best predictor skill for CT skills is, however, programming~\cite{kilicc2021valid}, but this isn't very helpful in the context of research where we try to find non-programming skills that predict software development with LLM-based tools. Interestingly, but again not very helpful, using ChatGPT to solve programming tasks is a good way to improve CT skills~\cite{yilmaz2023effect}. 

\subsection{Limitations}
This research has quite a few limitations, and part of the goal of this discussion paper is to present these limitations, and to discuss possible ways to address them.

First, there is no single category of programmers, or programmer tasks. Programmers can be compiler writers or front-end designers, and they can be working on a 5 year project with millions of lines of code or write formulas in an Excel spreadsheet. The latter category of programmers is included in the so-called ``end-user programmers''~\cite{scaffidi2005approach}. Scaffidi et al.~ estimated in 2005 that there were three to four times as many self-reported programmers than professional software developers, and almost twenty times as many spreadsheet and database users. The tasks these different kinds of programmers work on require sometimes similar, but sometimes also very different kinds of skills. In that sense the question we ask is not a good one: to determine what skills you need for programming, we first need to describe what \textit{kind} of programming we mean. We wrote this paper and asked our research question because of the very general statements made around the hype of using LLM-based tools when programming. We think our question addresses those statements, but we also think we should ask more precise questions. 

The second limitation of our work is that we tried to find out if \textit{CT skills} predict programming skills using LLM-based tools. This is based on the description of CT skills as skills a human needs to solve problems with the help of a computer. But wouldn't other skills, such as skills in mathematics or natural language~\cite{prat2020relating} better predict programming skills using LLM-based tools? Or wouldn't a subset of CT skills, such as decomposition of problems and evaluation of solutions be better predictors? 

The third limitation is that the CT skills test we used was validated for middle grades students, and not for adults. As CT skills tests go, the validation was thorough and extensive, but it might be the case that this test doesn't hold up for another age group.

In addition, we could have improved our experiment in several ways. We only set the participants a single task to solve using LLM-based tools. A training session together with setting multiple tasks with different characteristics might give different results, but would require a much larger time investment from our participants. Such a training session would also have mitigated the potentially confounding factor that some people with a programming background might have used ChatGPT for programming before our experiment, since ChatGPT had been out for some months at the time we performed our experiment. 

The participants used ChatGPT and an online interpreter to test the code generated by ChatGPT. Because the task asked for a user interface, testing the code was only possible for part of it. Using Github Copilot in Visual Studio Code might lead to different results. New versions of LLM-based tools will also likely give different results, but we expect the core of the ideas described in this paper remains the same. 

\section{Conclusions}
\label{sec:conclusions}
We have seen a flood of opinions on the possibilities and limitations of using LLM-based tools like ChatGPT and Github Copilot for programming in the last year. While it is interesting and useful to see examples of using LLM-based tools when programming, drawing conclusions like ``machines are going to program themselves'', the end of programming'', and ``AI does not help programmers'' are unfounded and unhelpful. Rather, we should investigate how the competencies of (categories of) programmers should evolve when developing software using LLM-based tools. Admittedly, the title of this paper contributes to the lack of nuance in the discussion around the use of AI in programming; it should really be something like: ``What skills does [a category of programmers] need when developing [a particular kind of] software using [a particular version of an LLM-based tool]''.  

In an experiment in which we let participants first work on a CT skills test, and then develop software using ChatGPT, we found that CT skills are a very good predictor for the ability to develop software using LLM-based tools. 
We found an almost perfect linear relation between the score on CT skills and the programming with ChatGPT score. Note that, of course, this correlation does not imply causation.

The limitations of this work lead to many interesting opportunities for future work. Testing which specific skills are needed for programming using LLM-based tools,
such as (1) program specification, (2) refactoring, and (3) verification, testing and evaluation, three of the new pedagogical approaches mentioned by Denny et al.~\cite{denny2023computing}, and predictors for these skills, is most likely going to lead to information about what we should teach our students. But how do we set up experiments to verify this? As for the participants, we should look at different groups. Skills for developing software using LLM-based tools are of course relevant to CS students, but given that there are many more end-user programmers with a different background, and that LLM-based tools are probably better at the kind of programs end-user programmers write, it is important to set up experiments with non-CS students. As for the experiment, we would need to design the instruments to measure both the predictors and the outcome, and preferably also an intervention to teach the development of software using LLM-based tools. There is a lot of interesting work ahead!\\

\noindent
\textbf{Acknowledgements.} This research was partially funded by the National Education Lab Artificial Intelligence through a grant of the National Growth Fund \#NGFNOLAI22.

    \balance
\bibliographystyle{ACM-Reference-Format}
\bibliography{bibliography}

\end{document}